\documentclass[3p,times,procedia,sort&compress]{elsarticle}

\usepackage{nupha_ecrc}

\usepackage[utf8]{inputenc}
\usepackage{amsmath}
\usepackage{wrapfig}

\volume{00}
\firstpage{1}
\journalname{Nuclear Physics A}
\runauth{J. E. Bernhard et al.}
\jid{nupha}
\jnltitlelogo{Nuclear Physics A}

\newcommand{\trento}{T\raisebox{-0.5ex}{R}ENTo}
\newcommand{\order}[1]{$\mathcal O(10^{#1})$}
\newcommand{\x}{\mathbf x}
\newcommand{\y}{\mathbf y}
\newcommand{\tran}{^\intercal}

\begin{document}

\begin{frontmatter}

\dochead{XXVIth International Conference on Ultrarelativistic Nucleus-Nucleus Collisions\\ (Quark Matter 2017)}

\title{Characterization of the initial state and QGP medium from a combined Bayesian analysis of LHC data at 2.76 and 5.02 TeV}

\author{Jonah E. Bernhard}
\ead{jeb65@phy.duke.edu}
\author{J. Scott Moreland}
\author{Steffen A. Bass}

\address{Department of Physics, Duke University, Durham, NC 27708}

\begin{abstract}
  We perform a global Bayesian analysis of a modern event-by-event heavy-ion collision model and LHC data at $\sqrt s = 2.76$ and 5.02 TeV.
  After calibration, the model simultaneously describes multiplicity, transverse momentum, and flow data at both beam energies.
  We report new constraints on the scaling of initial-state entropy deposition and QGP transport coefficients, including a quantitative estimate of the temperature-dependent shear viscosity $(\eta/s)(T)$.
\end{abstract}


\end{frontmatter}

\section{Introduction}

A primary goal of heavy-ion physics is the quantitative determination of the properties of the quark-gluon plasma (QGP), such as its transport coefficients and the characteristics of the initial state that leads to its formation.
Since the QGP medium created in ultra-relativistic heavy-ion collisions is highly transient, its properties are not directly measurable---but they may be estimated by comparing computational models to experimental observations.
The desired properties are input as model parameters and optimized so that the model's simulated observables best describe corresponding experimental data.
Previous studies have used Bayesian model-to-data comparison to place preliminary constraints on salient QGP properties such as the temperature dependence of the specific shear viscosity $(\eta/s)(T)$ and the scaling of initial-state entropy deposition \cite{Bernhard:2015hxa, Bernhard:2016tnd}.

In this work, we perform an improved Bayesian analysis of a heavy-ion collision model and experimental data from Pb+Pb collisions at $\sqrt s = 2.76$ and 5.02 TeV at the LHC.
We calibrate the model to multiplicity, transverse momentum, and flow data and report the latest quantitative estimates of the temperature dependence of QGP transport coefficients as well as initial state properties.

\section{Model}

Heavy-ion collision events are simulated using a modern multi-stage model with Monte Carlo event-by-event initial conditions, a pre-equilibrium free-streaming stage, viscous relativistic hydrodynamics, and a hadronic afterburner.

Initial conditions are generated by the parametric model \trento\ \cite{Moreland:2014oya, Bernhard:2016tnd}.
After sampling nucleon positions from a Woods-Saxon distribution and computing the participant nuclear thickness functions $T_A,T_B$, \trento\ deposits entropy according to the ansatz
\begin{equation}
  s \propto \Biggl( \frac{T_A^p + T_B^p}{2} \Biggr)^{1/p},
  \label{eq:trento}
\end{equation}
where $p$ is a continuous tunable parameter that effectively interpolates among different entropy deposition schemes.
When $p = 1$, the ansatz reduces to a wounded nucleon model ($s \propto T_A + T_B$), while $p = 0$ implies entropy deposition proportional to the geometric mean of thickness functions ($s \propto \sqrt{T_AT_B}$), which mimics successful saturation-based models such as IP-Glasma \cite{Schenke:2012wb} and EKRT \cite{Niemi:2015qia}.

Initial conditions are then free-streamed \cite{Liu:2015nwa} for a tunable time $\tau_\text{fs}$;
the energy density, flow velocity, and viscous pressures after free streaming serve as the complete initial condition for hydrodynamic evolution.

The hot and dense QGP medium is modeled by VISH2+1 \cite{Shen:2014vra, Bernhard:2016tnd}, an implementation of boost-invariant viscous relativistic hydrodynamics including temperature-dependent shear and bulk viscosities.
For the specific shear viscosity $\eta/s$, we use the modified linear ansatz
\begin{equation}
  (\eta/s)(T) = (\eta/s)_\text{min} + (\eta/s)_\text{slope}(T - T_c) \times (T/T_c)^{(\eta/s)_\text{curvature}},
  \label{eq:shear}
\end{equation}
where $\eta/s$ min, slope, and curvature are tunable parameters and $T_c = 0.154$ GeV is the equation of state transition temperature.
We parametrize the specific bulk viscosity $\zeta/s$ as a Cauchy distribution with tunable maximum and width, and peak location fixed at $T_c$:
\begin{equation}
  (\zeta/s)(T) = \frac{(\zeta/s)_\text{max}}{1 + \bigl[ (T - T_c)/(\zeta/s)_\text{width} \bigr]^2}.
  \label{eq:bulk}
\end{equation}
The hydrodynamic equation of state (EOS) consists of a hadron resonance gas EOS at low temperature connected to the HOTQCD lattice EOS \cite{Bazavov:2014pvz} at high temperature.

As the hydrodynamic medium expands and cools, it is converted to an ensemble of hadrons on an isothermal spacetime hypersurface defined by a tunable temperature $T_\text{switch}$.
Particle species and momenta are sampled from a thermal hadron resonance gas, including random Breit-Wigner masses for unstable resonances.
Shear and bulk viscous corrections are applied based on the relaxation-time approximation \cite{Pratt:2010jt, Dusling:2011fd}.

Finally, after the conversion to particles, the UrQMD model simulates the non-equilibrium expansion and breakup of the hadronic system.

\section{Parameter estimation}

\begin{figure}[t]
  \centering
  \includegraphics[width=.9\textwidth]{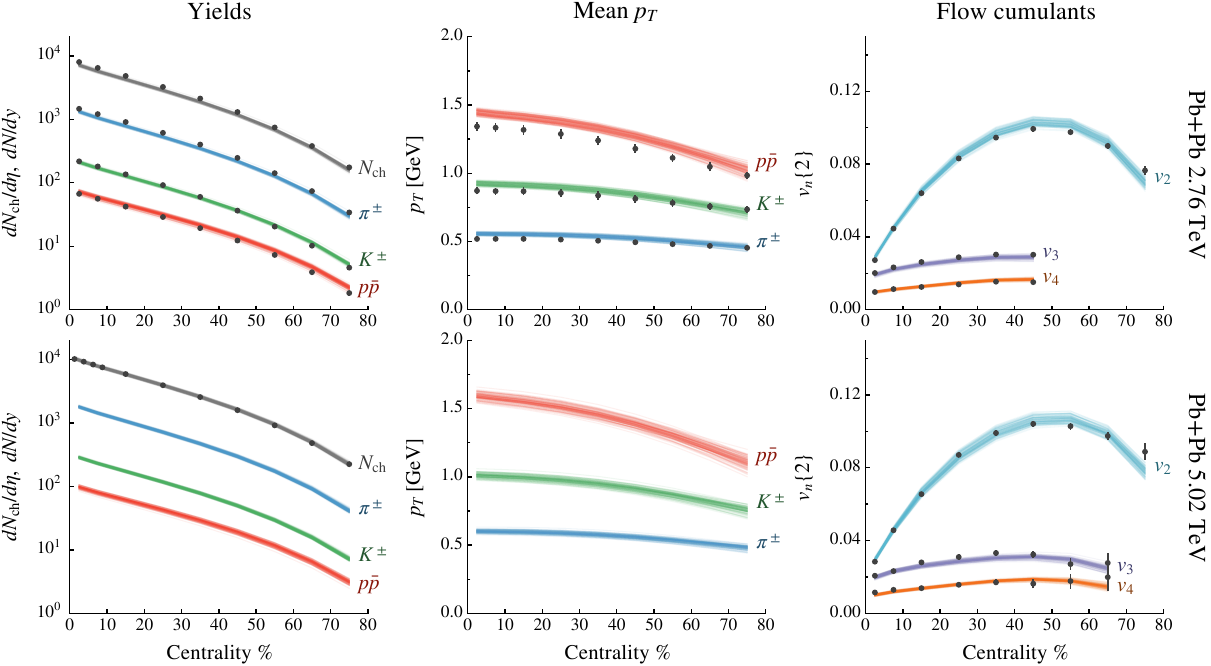}
  \caption{
    Model observables sampled from the posterior distribution (lines) compared to experimental data (points with error bars) \cite{Aamodt:2010cz, Abelev:2013vea, Adam:2015ptt, Adam:2016izf}.
    At the time of this writing, identified particle data were not available at $\sqrt s = 5.02$ TeV.
  }
  \label{fig:observables_posterior}
\end{figure}

The goal is now to calibrate the model to optimally describe experimental data and thereby extract a Bayesian posterior probability distribution for the true values of each model parameter.
Markov chain Monte Carlo (MCMC) methods can systematically explore the parameter space and produce the desired posterior distribution, but require millions of model evaluations.
In this case, a single model evaluation requires thousands of individual event simulations and hence thousands of computing hours, so direct MCMC sampling is intractable.
To circumvent this limitation, we utilize a modern Bayesian method for estimating the parameters of computationally expensive models \cite{Bernhard:2015hxa, Bernhard:2016tnd}, briefly summarized here.

We first evaluate the model at 500 parameter points chosen by Latin-hypercube sampling.
Each model evaluation consists of \order 4 minimum-bias events which are sorted into centrality bins and used to compute observables in analogy with experimental methods.
We compare to charged-particle yields, identified particle yields and mean transverse momenta, and azimuthal flow coefficients measured by the ALICE experiment at the LHC from Pb+Pb collisions at $\sqrt s = 2.76$ and 5.02 TeV \cite{Aamodt:2010cz, Abelev:2013vea, Adam:2015ptt, Adam:2016izf}.

We then interpolate the model using Gaussian process (GP) emulators;
given an arbitrary parameter vector $\x$, the GPs predict the corresponding model output $\y = \y(\x)$, including the uncertainty of the prediction.
The posterior probability at $\x$ is then
\begin{equation}
  P(\x) \propto \exp\bigl[ -\tfrac{1}{2} (\y - \y_\text{exp})\tran \Sigma^{-1} (\y - \y_\text{exp}) \bigr],
\end{equation}
where $\y_\text{exp}$ is the experimental data and $\Sigma$ is the covariance matrix, which is the total of experimental statistical and systematic uncertainty, model statistical uncertainty, and GP predictive uncertainty.

Finally, the posterior distribution is constructed by MCMC sampling, using the GPs as a fast surrogate for the full model.

\section{Results and discussion}

\begin{wrapfigure}{r}{.5\textwidth}
  \vspace{-1em}
  \includegraphics[width=.5\textwidth]{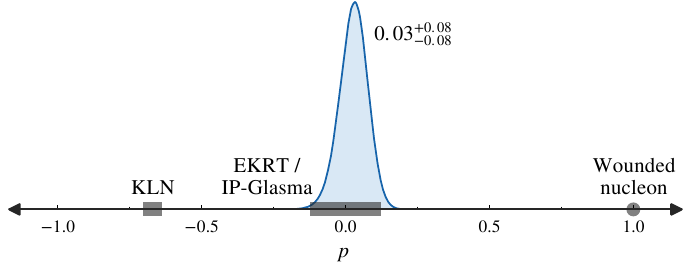}
  \caption{
    Posterior distribution of the \protect\trento\ entropy deposition parameter $p$ [defined in Eq.~\eqref{eq:trento}].
    The annotated value and uncertainty is the posterior median and 90\% credible interval.
    The approximate $p$-values of several existing initial condition models are labeled on the axis.
  }
  \label{fig:posterior_p}
\end{wrapfigure}

Figure~\ref{fig:observables_posterior} shows emulator predictions of model output sampled from the posterior distribution, compared to the experimental data.
We observe a good simultaneous to fit (within ${\sim}10\%$) to all data points at both beam energies.
The visual spread in the sample lines reflects the width of the posterior distribution, which arises from the various sources of experimental and model uncertainty as well as from tension between the model and data.

The full posterior distribution is beyond the scope of this proceedings;
we highlight only the few most important parameters.
First, the \trento\ entropy deposition parameter $p$ is shown in Fig.~\ref{fig:posterior_p}.
The distribution is strongly peaked near zero, implying that entropy deposition scales approximately as the geometric mean of local nuclear density ($s \approx \sqrt{T_AT_B}$).
This result has reduced uncertainty compared to previous work and corroborates saturation-based models such as IP-Glasma \cite{Schenke:2012wb} and EKRT \cite{Niemi:2015qia}.

\begin{figure}
  \includegraphics[height=.37\textwidth]{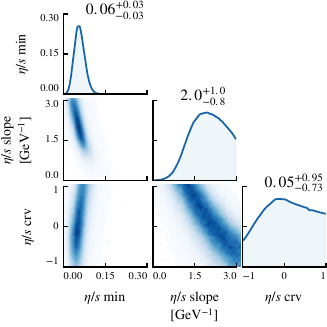}
  \hfill
  \includegraphics[height=.37\textwidth]{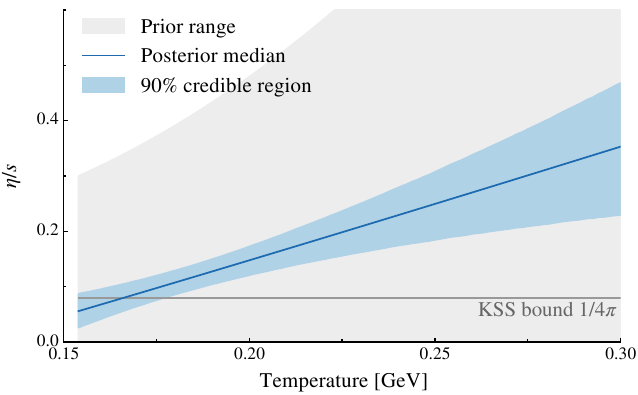}
  \caption{
    Left: Posterior distribution of the $(\eta/s)(T)$ parameters [defined in Eq.~\eqref{eq:shear}].
    The annotated values and uncertainties are the posterior medians and 90\% credible intervals.
    Right: Visualization of the estimated temperature dependence of $\eta/s$.
  }
  \label{fig:shear}
\end{figure}

Figure~\ref{fig:shear} shows the shear viscosity parameters and the corresponding estimate of the temperature-dependent curve $(\eta/s)(T)$.
The distribution for $(\eta/s)_\text{min}$ (the value of $\eta/s$ at the transition temperature $T_c = 0.154$ GeV) has a narrow peak at 0.06, below the KSS bound $1/4\pi \approx 0.08$ but consistent within 90\% uncertainty.
On the other hand, zero $\eta/s$ is excluded at the 90\% level.
The slope parameter has a broad peak, although zero slope is excluded, thus confirming a temperature-dependent (non-constant) $\eta/s$ at the 90\% level.
The curvature parameter is not constrained, but exhibits a strong correlation with the slope.
In the right panel of Fig.~\ref{fig:shear}, we visualize the estimated temperature dependence of $\eta/s$ by inserting the posterior samples for the $\eta/s$ parameters back into Eq.~\ref{eq:shear}.
We emphasize that the pronounced narrowing of the uncertainty band at low temperature was not assumed---it is a natural consequence of the resolving power of the data.
Including data from additional beam energies (i.e.\ RHIC) could reduce the uncertainty on $(\eta/s)(T)$ and possibly constrain the curvature in addition to the minimum and slope.

The posterior distribution for the bulk viscosity parameters $\zeta/s$ max and width [see Eq.~\eqref{eq:bulk}] shows that $(\zeta/s)(T)$ may be ``tall'' [$(\zeta/s)_\text{max} \gtrsim 0.02$] or ``wide'' [$(\zeta/s)_\text{width} \gtrsim 0.01$ GeV], but not both.
However, this result is sensitive to the precise implementation of bulk viscous corrections at particlization and is therefore subject to change.

\vspace{.5em}\noindent\textbf{Acknowledgments}:
The Duke QCD group acknowledges support by grants no.\ NSF-ACI-1550225 (NSF) and DE-FG02-05ER41367 (DOE).
CPU time was provided by the Open Science Grid, supported by DOE and NSF, as well as the DOE funded National Energy Research Scientific Computing Center (NERSC).

\bibliographystyle{elsarticle-num}
\bibliography{proceedings}

\end{document}